\documentstyle[sprocl]{article}

\input{psfig.sty}

\def\be{\begin{eqnarray}}
\def\ee{\end{eqnarray}}

\newcommand{\mat}{\left ( \begin{array}{cc}}
\newcommand{\emat}{\end{array} \right )}
\newcommand{\matt}{\left ( \begin{array}{ccc}}
\newcommand{\ematt}{\end{array} \right )}
\newcommand{\matf}{\left ( \begin{array}{cccc}}
\newcommand{\ematf}{\end{array} \right )}
\newcommand{\vect}{\left ( \begin{array}{c}}
\newcommand{\evect}{\end{array} \right )}

\begin{document}

\title{DIRAC SPECTRUM IN ADJOINT QCD}

\author{D. TOUBLAN AND J.J.M. VERBAARSCHOT}

\address{Department of Physics and Astronomy,\\
         University at Stony Brook,\\
         Stony Brook, NY\,11794, USA}   


\maketitle\abstracts{In this lecture we discuss some exact results
for the low-lying spectrum of the Dirac operator in adjoint QCD. In particular,
we find an analytical expression for the slope of the average spectral density.
These results are obtained by means of 
 a generating function which is  an 
extension of the QCD partition function with fermionic and bosonic ghost quarks.
The low-energy limit of this generating function is completely determined
by chiral (super-)symmetries. Our results for the slope of the average
spectral density are
consistent with the results for the scalar susceptibility which can be
obtained from the usual chiral Lagrangian.}

\section{INTRODUCTION}
Both from phenomenological arguments and lattice QCD simulations we
know that chiral symmetry in QCD is spontaneously broken by the
formation of a chiral condensate 
(this issue has been discussed in several recent reviews 
\cite{DeTar,Smilref,Kyoto,Tilorev,Creutz}). However, a complete analytical 
understanding of the underlying 
mechanism of chiral symmetry breaking is not yet available. The
situation is much better in Supersymmetric Gluodynamics. In this theory
it can be shown analytically that the chiral condensate is
non-vanishing \cite{NSVZ,Davies1999,Ritz,Eric}. One important difference
with QCD is that in this case the fermions are in the adjoint 
representation. 
In both cases the partition function can be viewed
as the average of a fermion determinant. The chiral condensate, which
is the mass derivative of the free energy, is thus directly related
to the eigenvalues of the Dirac operator. 

Generally, the Dirac
spectrum cannot be obtained analytically. However, because the
low-energy limit of theories with Goldstone bosons and a mass gap
is uniquely determined by the pattern of spontaneous
symmetry breaking, we expect that we will be able to derive analytical results
for the low-lying Dirac spectrum. This program was initiated by 
Leutwyler and Smilga \cite{LS} whose work resulted in sum-rules for
the inverse Dirac eigenvalues. A complete analytical understanding
of the low-lying Dirac spectrum came from the
realization that it is described by a Random Matrix Theory with
the global symmetries of QCD, also known as chiral Random Matrix Theory 
\cite{SV,V}. 
This conjecture has been proved analytically starting from the low-energy
limit of a generating function for the Dirac spectrum \cite{OTV,DOTV}.
In addition to the usual fermionic quarks, this QCD-like partition function
contains fermionic and bosonic ghost quarks with a mass 
determined by the magnitude of the Dirac eigenvalues we are interested in.
It was understood early on \cite{Vplb,james} that the domain of 
validity of chiral 
Random Matrix Theory is determined by the mass scale for which the 
kinetic term of this low-energy effective theory can be neglected, i.e. when
the wavelength of the corresponding Goldstone modes is much larger than the
size of the box. In this domain, the thermodynamic limit can only be
taken if the mass is decreased such that this condition remains
satisfied. However, this is exactly what happens
for the low-lying Dirac eigenvalues which scale as the inverse
Euclidean volume. 
The recent work on the generating function for the Dirac spectrum 
\cite{OTV,DOTV,TV} made it possible to include the effect of the
kinetic term and to study the Dirac spectrum in the physical domain with
box size much larger than the Compton wavelength of the Goldstone modes.
This allowed us to extract properties of the average spectral density
at a scale that remains fixed
 in the thermodynamic limit. In particular, an analytical
expression for  its  slope was found \cite{OTV,TV}.

In this lecture we are interested in the QCD Dirac spectrum for
fermions in the adjoint representation. We will calculate the slope
of the Dirac spectral density in two different ways. First, via the scalar
susceptibility which can be calculated by means of the usual chiral
Lagrangian, and second, via the valence quark mass dependence
of the chiral condensate. The first method was originally introduced
by Smilga and Stern \cite{Stern} for the case of  QCD with three or more
colors and fundamental fermions.
The second approach relies on the 
introduction of a generating function for the resolvent of the 
QCD Dirac operator as discussed above.
We will find that both methods give the same
results. One of the advantages of the second method, which proceeds
by a direct calculation of the average spectral density of the Dirac
operator, is its validity for $N_f =1$.

In our
convention, with an anti-Hermitian Dirac operator $D$, the  eigenvalues
are given by
\be
D\phi_k = i\lambda_k \phi_k.
\ee
The average spectral density is defined as
\be
\rho (\lambda) = \langle \sum_k \delta(\lambda -\lambda_k)\rangle,
\ee
where the average is over the ensemble of spectra.
There are two important  observables that are directly related
to the Dirac spectrum, the chiral condensate and the scalar susceptibility.
The chiral condensate is given by,
\be
\Sigma &\equiv& 
\lim_{m\to 0} \lim_{ V\to \infty} \frac 1V \int \frac {\rho(\lambda) 
d\lambda}{i\lambda +m},
\ee
and the scalar susceptibility can be written as, 
\be
K & =& 
-\lim_{V\to \infty}
\frac 1{V} \int \frac{\rho(\lambda) d \lambda}{(i\lambda + m)^2}.
\ee
Here, and below, the Euclidean space time volume is denoted by $V$. 
If the spectral density near zero can be expanded as 
\be
\rho(\lambda) = \rho(0) + |\lambda |\rho'(0) +
 \frac 12 \rho''(0) \lambda^2
\cdots,
\label{rholin}
\ee
the chiral condensate is given by by the Banks-Casher formula \cite{BC},
\be
\Sigma = \frac {\pi \rho(0)}{V},
\ee
and the infrared singular part of the scalar susceptibility is given by
\be
K \sim \frac{\rho'(0)}V \log(\Lambda/m).
\ee
The chiral condensate is obtained from the first term in
(\ref{rholin}). This term does not contribute to $K$; the infrared
singular part arises from the linear term in (\ref{rholin}).
All higher order terms in (\ref{rholin}) can be neglected
if the chiral limit is taken at fixed
value of the cutoff $\Lambda$ for the integration over $\lambda$.
The smallest Dirac eigenvalues thus provide us with important
information about the vacuum properties of QCD.

\section{QCD WITH ADJOINT FERMIONS}

The Euclidean Dirac operator for quarks in the adjoint representation 
is given by
\be
D = \gamma_\mu(\partial_\mu \delta_{bc} + f_{abc} A^a_\mu),
\ee
where the $f_{abc}$ are the anti-symmetric structure constants of $SU(N_c)$ 
and the $\gamma_\mu$ are the Euclidean $\gamma$-matrices. Because the gauge
fields $A^a_\mu$ are real this Dirac operator satisfies the reality
relation \cite{Teper,LS,V}
\be
[iD,CK] = 0,
\label{reality}
\ee 
where $C$ is the charge conjugation matrix ($C= \gamma_2\gamma_4$) 
and $K$
is the complex conjugation operator. Because
\be
(CK)^2 =-1,
\ee
all Dirac eigenvalues are doubly degenerate with eigenfunctions given
by $\phi$ and $CK \phi$. The linear independence of $\phi$ and $CK\phi$
follows \cite{LS} from properties of the scalar product under anti-unitary
transformations,
\be
(CK\phi, \phi) = ((CK)^2 \phi, CK \phi)^* = -(CK\phi, \phi), 
\ee
so that $(CK\phi, \phi)= 0$. Another consequence of (\ref{reality}) is
that it is always possible to find a basis for which the 
matrix elements of the Dirac operator are arranged into self-dual
quaternions \cite{HVeff}. In other words, the Dyson index of the
Dirac operator in adjoint QCD has the value $\beta = 4$.

Because $\gamma_2\gamma_4(D+m)$ is anti-symmetric under transposition, the
square root of the fermion determinant is given by its Pfaffian, and the 
Euclidean partition function for $N_f$ Majorana flavors can be written
as
\be
Z = \int D A \, {\det}^{1/2}(D +M) e^{-S_{\rm YM}}=
\int D A \int \prod_{f=1}^{N_f} d\lambda^f 
e^{ \int d^4 x \lambda^f C (D \delta_{fg} +M_{fg}) \lambda^g} 
e^{-S_{\rm YM}}.\nonumber \\
\ee
The mass matrix $M$ is symmetric under transposition.
In the case  of one massless flavor 
this theory is Supersymmetric Gluodynamics. Its
properties have been investigated in great detail 
(an excellent introductory review of this topic is available \cite{Shifman}).
In particular, it has been shown \cite{NSVZ} that the gluino
condensate is non-vanishing. Since the adjoint Dirac operator in the the field
of an instanton has $2N_c$ zero modes, this result cannot be understood
in terms  of explicit symmetry breaking by instantons as is the case 
for QCD with
one massless fundamental flavor.
A non-vanishing chiral condensate
in Supersymmetric Gluodynamics thus suggests the existence 
of field configurations with winding number equal to $1/N_c$.
Considerable progress was made in this direction in several recent
articles \cite{Eric,Davies1999,Kraan,Lee}. Some evidence for the
existence of such configurations has been found in lattice QCD as well
\cite{hellerad}.

\section{DIRAC SPECTRUM }

In this section we review some properties of the QCD Dirac spectrum. 
Because of the axial $U_A(1)$ symmetry, $\{D, \gamma_5\} =0 $, 
all nonzero eigenvalues occur in pairs
$\pm \lambda_k$ with eigenfunctions given by $\phi_k$ and $\gamma_5 \phi_k$.
If $\phi_k \sim \gamma_5 \phi_k$, the corresponding eigenvalue is 
necessarily zero. This happens in the field of an instanton. 

For eigenvalues much larger than $\Lambda_{\rm QCD}$, 
we expect that the gauge fields do not 
significantly modify the Dirac spectrum so that its spectral density 
is given by a theory of noninteracting quarks,
\be
\rho(\lambda) \sim V \lambda^3 \quad {\rm for } \quad \lambda \to \infty.
\ee

\begin{center}
\begin{figure}[!ht]
\hspace{1.5cm}
\hbox{\psfig{file=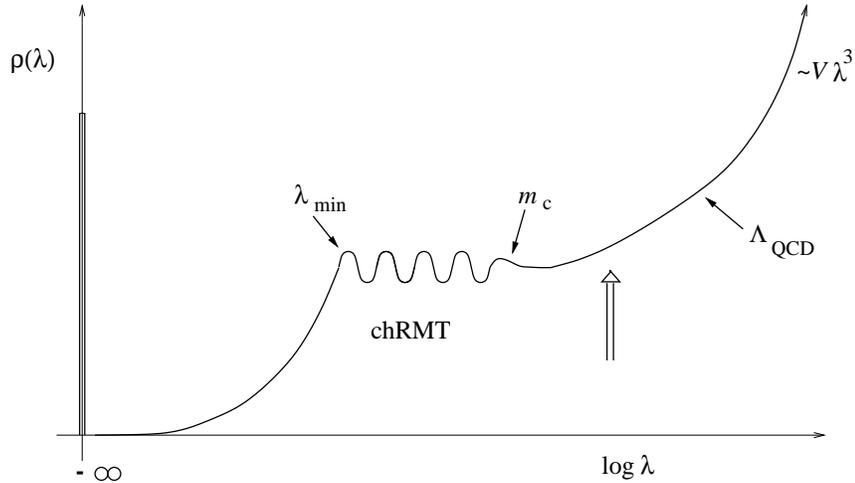,height=64mm}}
\caption[]{Schematic picture of the QCD Dirac spectrum. The arrow
denotes the region of interest of the present article. }
\label{fig1}
\end{figure}
\end{center}
In the case of spontaneous broken chiral symmetry, the chiral condensate
is nonzero if the thermodynamic limit is taken before the chiral
limit. This can be understood in terms of the existence of a tower of
roughly equally spaced eigenvalues  (indicated by the wavy curve
in Fig. 1) with the smallest 
nonzero eigenvalue at about one average spacing from zero,
\be
\lambda_{\rm min} \approx \frac 1{\rho(0)} = \frac {\pi}{\Sigma V}.
\ee

Such accumulation of eigenvalues near zero does not occur in the free
theory. It is only possible if the Dirac spectrum near
zero is dominated by the interactions of the theory. Strong
interactions give rise to repulsion of the eigenvalues which,
viewed as positions of particles, condense into a Wigner crystal.
This phenomenon has been studied in great detail in
the context of Random Matrix Theory \cite{Tilorev,HDgang}. 
As will be explained next, 
the smallest QCD Dirac eigenvalues are correlated according
to chiral Random Matrix Theory (chRMT) \cite{SV,V}.

An important energy scale in the Dirac spectrum is the Thouless energy
\cite{GL,LS,Vplb,james}. This is the quark 
mass scale $m_c$,  for which the Compton wavelength of the corresponding
Goldstone boson is equal to the length of the box, i.e.
\be
\frac{m_c \Sigma}{F^2} = L^2.
\ee
For $m \ll m_c$ the kinetic term of the Goldstone modes can be
ignored.  In this domain, all theories with the same pattern of chiral
symmetry breaking and a mass gap are equivalent (several explicit examples
have been constructed \cite{Gade,Simons,Taka,TTH}). In particular, 
the Dirac spectrum  below $m_c$ is
 given by a chiral Random Matrix Theory \cite{OTV,DOTV}. 
For adjoint fermions this is
a random Dirac operator with quaternion real matrix elements with 
a probability distribution that includes the fermion determinant
\cite{V}.
This has been confirmed
by numerous lattice QCD simulations 
\cite{HV,Vplb,berbenni,many,Damg99a,TiloSU3,Edwa99b,Berg99,Fa,Berb99,Damhnr}; 
(a complete list of references can be found in a recent 
review \cite{Tilorev}). 
The basis for the predictive power of
Random Matrix Theory is universality: the fluctuation properties
of the eigenvalues on the scale of the average level spacing are
not sensitive to a wide class of large modifications of the 
probability distribution. In chiral Random Matrix Theory, this was
first shown for QCD with three or more colors and fundamental fermions
\cite{brezin,ADMN} and only more recently for the case of adjoint fermions
\cite{Senerprl,Klein,Wido98,akekan,Damg98a}.

Should we also expect an accumulation of small Dirac eigenvalues for one
massless Majorana flavor? 
Let us consider
the chiral condensate,
\be
\langle \bar \lambda \lambda \rangle = \frac 1Z \frac 1V
\left \langle\sum_k \frac 1{ i\lambda_k + m} \prod(i\lambda_k +m)
\right \rangle.
\label{ll}
\ee
For $ m \to 0$ the partition function is dominated by configurations
with zero topological charge, whereas the numerator in (\ref{ll}) 
obtains its
main contribution from the $\nu =1/N_c\equiv\bar \nu/N_c$ 
configurations for which the sum over
eigenvalues in (\ref{ll}) 
can be approximated by the $\lambda_k = 0$ term. 
With a cancellation of the $1/m$ factor from the chiral condensate and 
a factor $m$ in the fermion determinant, we thus find
\be
\langle \bar \lambda \lambda \rangle = \frac 1V
  \frac{\langle \prod_k i\lambda_k \rangle_{\bar \nu = 1}}
  {\langle \prod_k i\lambda_k \rangle_{\bar \nu = 0}}.
 \ee
The average fermion determinants for $\bar \nu =0$ and $\bar\nu = 1$ differ
by a factor  $1/\langle \bar \lambda \lambda \rangle V$. How can we 
understand this? Our explanation is that the zero eigenvalue
repels all nonzero eigenvalues such that their average position
(denoted by $\bar \lambda_k$)
is shifted away from
zero by exactly one half average eigenvalue spacing $\Delta \lambda$. 
We thus have
\be
\prod (\bar \lambda_k^{\bar \nu = 1})^2 \approx  
\prod (\bar\lambda_k^{\bar \nu = 0 } + \frac 12 \Delta \lambda)^2 
\approx  \prod \bar\lambda_k^{\bar \nu = 0 }
\bar\lambda_{k+1}^{\bar \nu = 0 } = 
\lambda_{\rm min} \prod (\bar \lambda_k^{\bar\nu = 1})^2.
\ee 
Explicit chiral symmetry breaking by instantons
also requires the accumulation of 
small nonzero eigenvalues exactly as happens in the case of spontaneous
breaking of chiral symmetry when the chiral condensate is given by the
Banks-Casher formula \cite{BC}. These results are in agreement with the
analysis based on finite volume partition functions for $N_f =1$ for which
the contributions to the chiral condensate 
from the different topological sectors
can be obtained analytically \cite{LS,Poultop}.

\section{CHIRAL LAGRANGIAN AND SCALAR SUSCEPTIBILITY}
The adjoint QCD partition function is invariant under
\be
\lambda \equiv \vect w \\ \bar w \evect \to \mat U & \\ & U^{-1} \emat 
\vect w \\ \bar w \evect,
\ee
where  $U \in SU(N_f)$ (a $U_A(1)$ axial symmetry is broken by the anomaly).
The gluino condensate is a color singlet with flavor structure 
given by \cite{SmV}
\be
\langle \lambda^f C \lambda^g \rangle = \delta^{fg}\, \Sigma,
\label{condensfg}
\ee
so that the $SU(N_f)$ flavor symmetry is broken to $O(N_f)$. The chiral
Lagrangian corresponding to this pattern of chiral symmetry breaking can 
be constructed in the standard way. To lowest order in the momenta and the
quark masses it is given by \cite{TV}
\be
Z(M) = \int_{U\in SU(N_f)/O(N_f)} dU
e^{-\int d^4 x [\frac {F^2} 4 {\rm Tr} \partial_\mu U \partial_\mu U^{-1}
- \frac{\Sigma}2 {\rm Tr} (MU +M^\dagger U^{-1})]}.
\ee
This partition function is invariant under the flavor transformations
\be
U \to V U V^T, \qquad M \to V^* M V^\dagger, 
\ee
as required by the transformation properties of the QCD partition function.
Such invariance arguments are very powerful and even allow us to
construct the low energy limit of adjoint 
QCD at nonzero baryon density \cite{kstvz}, but this topic will not be
addressed today.

To calculate the scalar susceptibility it is convenient to introduce
scalar sources in the mass matrix of the QCD partition function
\be
M = m\,\delta^{fg} + s_a T^a,
\ee 
where the $T_a$ are the symmetric generators of $SU(N_f)$. The 
scalar susceptibility is then given by  the second derivative of the
QCD partition function with respect to the scalar sources
\be
K_{ab} &=& \frac 1{V } \partial_{s_a} \partial_{s_b} 
\bigr |_{s_a= 0, s_b = 0} \log Z(M),
\nonumber \\
& = & -\frac 1{V} \left \langle {\rm Tr} \frac {T_a}{D+m}\frac {T_b}{D+m} 
\right \rangle = \frac 12\delta_{ab} K.
\ee
Since the scalar sources and the mass matrix have the same
transformation properties in the QCD Lagrangian, they also enter in the
same way in the chiral Lagrangian.
To lowest order in chiral perturbation theory, 
we expand $U = \exp[ i2 \pi_a T^a/ F]$ to
second order in the $\pi$ fields,
\be
\frac 12(U + U^{-1}) = 1 - \frac 2{F^2} \pi_k \pi_l T^k T^l,
\ee
so that the scalar susceptibility is given by
\be
K^{ab} &=& \frac {4 \Sigma^2}{F^4} {\rm Tr }(T^a T^k T^l)
{\rm Tr }(T^b T^m T^n)
\langle \frac 1V
\int d^4 x d^4 y \pi_k(x) \pi_l(x) \pi_m(y) \pi_n(y) \rangle_{1-{\rm
loop}}\nonumber \\
&=& \frac{\Sigma^2}{16\pi^2 F^4}
{\rm Tr }(T^a \{T^k, T^l\}){\rm Tr }(T^b \{T^k, T^l\})\log (\Lambda/m).
\ee 
Carrying out the traces
in flavor space we find \cite{TV}
\be
K^{ab} = -\delta^{ab}\frac{\Sigma^2}{128 \pi^2 F^4}
\frac {(N_f-2)(N_f+4)}{N_f} \log
(m/\Lambda).
\label{kab}
\ee
The scalar susceptibility can also
be calculated for QCD with fundamental fermions and
two colors or for QCD with fundamental fermions and three
of more colors with Goldstone manifold given
by $SU(2N_f)/Sp(2N_f)$ and $SU(N_f)$, respectively. 
These three cases can be distinguished by the value of the Dyson
index given by  $\beta =4$, $\beta=1$ and $\beta=2$, in this order.
For Dyson index different from $\beta = 4$, the only change in (\ref{kab})
is the replacement $(N_f+4)/4 \to (N_f+\beta)/\beta$. 
The case $\beta = 2$ was first
analyzed by Smilga and Stern \cite{Stern}.

If the average spectral density has
a simple linear expansion as given in (\ref{rholin}), the slope
of the average spectral density follows immediately from the result
for the scalar susceptibility. It is given by
\be
\frac{\langle \rho'(0)\rangle }V &=&\Sigma^2
\frac{ (N_f-2) (N_f+\beta) }{16 \pi^2 \beta N_f F^4 },
\label{rhoQCD}
\ee
where we have included the dependence on the Dyson index.
In the next section we will present a derivation for the slope that
does not rely on this assumption (\ref{rholin}), and moreover, 
is also valid for $N_f =1$.

\section{GENERATING FUNCTION FOR THE DIRAC SPECTRUM}

In order to obtain a generating function for the Dirac spectrum 
one has to extend the QCD partition
function with additional fermionic and bosonic ghost quarks \cite{OTV},
\be
Z^{\rm pq}(z,J)  ~=~ \int\! DA
~\frac{\det(D +z+J)}{\det(D +z)}
{\det}^{N_f/2}(D + m) ~e^{-S_{YM}[A]} ~.
\label{zgen}
\ee
The resolvent is then given by
\be
\Sigma(z) = \frac 1V \left \langle \sum_k \frac 1{z +i\lambda_k} 
\right \rangle  =
\frac 1V \left . \partial_J \right |_{J=0} \log Z^{\rm pq} (z, J).
\label{resolv}
\ee
 The average spectral density follows from
the  discontinuity of the resolvent across the imaginary axis
\be
\frac{\rho(\lambda)}V &=& \frac 1{2\pi} (\Sigma (i\lambda +\epsilon) -
\Sigma( i\lambda - \epsilon)) \nonumber \\
&=&  \frac 1{2\pi} (\Sigma (i\lambda +\epsilon) +
\Sigma( -i\lambda + \epsilon)), \label{disc}
\ee
where the second equality follows from the relation 
$\Sigma(z)=-\Sigma(-z)$.
Notice that ${\det}^{-1/2} (D+z)$ cannot be written as a Gaussian integral.
The minimal generating function thus requires the introduction of  
ghost determinants as in eq. (\ref{zgen}), corresponding to one complex
bosonic ghost quark and a pair of Majorana ghost quarks.

In the sector of fermionic quarks, the symmetry is broken by the formation of
a chiral condensate with flavor structure as given in (\ref{condensfg}).
The pattern of symmetry breaking is thus given by
\be
SU(N_f +2) \to O(N_f +2).
\ee
In the sector of bosonic quarks, the quadratic form in the action can be
written as
\be
\vect \phi_L^* \\ \phi_R^* \evect 
\mat m &\sigma_\mu d_\mu \\ \sigma_\mu^\dagger d_\mu & m \emat 
\vect \phi_R\\  \phi_L \evect &=&
\vect \phi_L^*\\  \sigma_2 \phi_R \evect 
\mat   \sigma_\mu d_\mu &  \\ & \sigma_\mu d_\mu  \emat 
\vect \phi_L\\ \sigma_2\phi_R^*   \evect \nonumber \\ &+&
\frac 12 \vect  \phi_L^*\\  \sigma_2 \phi_R \evect 
\mat & m\sigma_2  \\ -m\sigma_2 &  \emat 
\vect \phi_L^* \\ \sigma_2\phi_R   \evect \\ &+&
\frac 12 \vect \phi_L \\  \sigma_2 \phi_R^* \evect 
\mat & m\sigma_2  \\ -m\sigma_2 &  \emat 
\vect \phi_L \\ \sigma_2\phi_R^*   \evect . \nonumber
\ee
Here, $\sigma_\mu = (1,i\sigma_k)$ and 
$d_\mu = \partial_\mu + f_{abc} A^a_\mu$.
The kinetic term is invariant under $U(2)$ transformations. However,
the fields no longer occur in complex conjugated pairs after this
transformation.  But notice that
the kinetic term is also invariant under the
symmetry group $U^*(2,R)$, which does not affect the reality
properties of the quadratic form. The chiral condensate given by the 
mass derivative of the partition function as well as the mass
term in the partition function are only left invariant 
by the subgroup $Sp(2)$. The Goldstone manifold corresponding to
 the sector of bosonic quarks 
is thus given by $U^*(2)/Sp(2)$ and has only 
one degree of freedom \cite{class,TV}.
  
In the sector of fermionic quarks we could extended the unitary
symmetry to $Gl(N_f+2)$, 
but a noncompact symmetry group would lead to an effective partition 
function with an incorrect small mass expansion. The symmetry group
of the sector of fermionic quarks should thus be $U(N_f+2)$.
In addition, the generating function is invariant under supersymmetry 
transformations mixing fermions and bosons.  The full symmetry group
is thus given by the graded Lie group $Gl(N_f+2|2) $ which is broken
spontaneously to the ortho-symplectic graded Lie group $Osp(2|N_f+2)$. The
Goldstone manifold is then given by \cite{class}
the maximum Riemannian submanifold
of $Gl(N_f+2|2)/Osp(2|N_f+2)$ with fermionic sector given by $U(N_f
+2)/O(N_f +2)$ and bosonic sector  $U^*(2)/Sp(2)$. We denote this
manifold by ${\widehat{G/H}} $.

 The low-energy effective partition function is given by 
\be
Z_{\rm eff} = \int_{U\in \widehat{G/H}} d U e^{-\int d^4 x {\cal L}_{\rm eff}},
\label{effpart}
\ee
where
\be
{\cal L}_{\rm eff}=\frac{F^2}{4} {\rm Str}(\partial_\mu U
\partial_\mu U^{-1})
- \frac{\Sigma}{2} {\rm Str}({\cal M} (U+ U^{-1}))
+{m_0^2} \Phi_0^2
+{\alpha} \partial_\mu \Phi_0 \partial_\mu \Phi_0
\label{Leff},\nonumber \\
\ee
and $U =\exp(i 2 \Phi/F)$.
The last two terms in (\ref{Leff}) represent the mass term and 
the kinetic term of the
super-$\eta'$ flavor-singlet field $\Phi_0={\rm Str}(\Phi)$. This partition
function has the same transformation properties as the generating
function (\ref{zgen})
including the explicit breaking of an axial $Gl(1|1)$ symmetry.
The mass matrix is given by ${\cal M} = {\rm diag} (m, \cdots, m, 
z+J,z+J,z,z)$, with $N_f$ masses equal to $m$. This partition function
has both bosonic and fermionic Goldstone bosons with mass  $M_{vv}= \sqrt
{2z\Sigma}/F$, 
$M_{vs}=\sqrt{(m+z)\Sigma}/F$ and $M_{ss}=\sqrt{2m\Sigma}/F$.
The generating function (\ref{zgen}) was first introduced to study the quenched
approximation in QCD \cite{Morel}. For symmetry class $\beta =2$ a 
 version of the effective
partition function based on compact supergroups
was first introduced by Bernard and Golterman \cite{pqChPT}. It gives
the correct perturbative expansion, but the nonperturbative
integrations over $U$ are not reproduced correctly.
Notice that a function is not determined by its asymptotic expansion.

To lowest order in chiral perturbation theory, the resolvent is simply 
given by  tadpoles  coming from the differentiation with respect
to the source $J$. There are three different types of contributions
corresponding to three different kinds of mesons that can be 
excited by the source $J$. First,
there are tadpoles with bosonic mesons of mass $M_{vs}$ that do not mix with
the super-$\eta '$, second, 
there are  tadpoles with fermionic mesons of mass $M_{vv}$, 
and third there are tadpoles with bosonic mesons of mass
$M_{vv}$ that mix with the super-$\eta '$. 
We thus find the following result for the resolvent (\ref{resolv})
\be
\Sigma(z)&=&
\Sigma_0\left [ 1 - \frac 1{F^2} \left \{ \frac{N_f}2 \Delta(M_{vs}^2)
 - \frac 12\Delta(M_{vv}^2) +G_{vv}\right \} \right ],
\ee
where the three terms in-between the braces correspond to the three
different types of  contributions discussed above, respectively. 
The trace of the propagator
of the first two types of mesons is given by
\be 
\Delta(M^2) = \frac 1V \sum_p \frac 1{p^2 +M^2}=\frac 1{16\pi^2} M^2 \log 
\frac {M^2}{\Lambda^2},
\ee
where $\Lambda$ is a momentum cutoff. 
The propagator of the third type of mesons is more complicated 
but it is known
analytically \cite{TV}. In the limit of $m_0 \to \infty$,
the trace of this propagator simplifies to
\be
G_{vv} = \frac 1V \sum_p\left [ \frac 1{p^2+M^2_{vv}} -\frac 1{N_f} 
\frac {p^2 +M_{ss}^2}{(p^2+M_{vv}^2)^2}\right ].
\ee
 
Using the explicit expressions for the trace of the propagators we find,
 \be
\Sigma(z)=\Sigma \; \Bigg[ 1- \frac{\Sigma}{16 \pi^2 N_f F^4}
\Bigg\{ \frac{N_f^2}2 (z+m) \log \frac{z+m}{2 \mu}
+\Big( 2 m+ (N_f-4)  z \Big)
\log \frac{z}{\mu} \Bigg\} \Bigg], \nonumber \\
\label{sigmaT}
\ee
where $\mu=\Lambda^2 F^2/2 \Sigma$.
In the limit of $m \rightarrow 0$ this result simplifies to \cite{TV}
\be
\Sigma(z)=\Sigma \; \Bigg[ 1- \frac{\Sigma  (N_f-2)(N_f +4)}{32 \pi^2 N_f F^4}
z \log \frac{z}{\mu}
\Bigg]\label{sigmaTmsz}.
\ee

~From the discontinuity of the resolvent along the imaginary axis 
calculated using the second equation of (\ref{disc})
 we find that the spectral density  in the limit $m \to 0$ is 
given by \cite{OTV,TV}
\be
\frac{\langle \rho(\lambda)\rangle}V &=&\frac{\Sigma}{\pi}
\Bigg[ 1+\frac{ (N_f-2) (N_f+\beta) \Sigma}{16 \pi \beta N_f F^4 }
|\lambda|\Bigg],
\label{rhoQCD2}
\ee
where the Dyson index is $\beta = 4$ in the case of adjoint QCD. 
We have also  included the results for QCD with two colors and
fundamental fermions ($\beta = 1$) and QCD with three or more colors
and fundamental fermions ($\beta =2$). The latter two cases  can be
derived along the same lines \cite{OTV,TV}. This result is in
agreement with the slope obtained from the scalar susceptibility. It
shows that the spectral density can be expanded in powers of $|\lambda|$.
Finally, we wish to emphasize that the above derivation is also valid
for $N_f = 1$. Results from instanton liquid simulations
\cite{Vinst,Thomas} are consistent with eq. (\ref{rhoQCD2}).  
 
As an alternative to the supersymmetric method, the mass dependence of
the resolvent can be calculated by introducing $n$ flavors of
fermionic ghost quarks with mass $z$ and take the limit $n \to 0$ at
the end of the calculation. This so-called replica method was used 
to derive the low-energy limit of the quenched
scalar susceptibility in lattice QCD with staggered fermions \cite{Berb99}.
A critical comparison of the supersymmetric
calculation and the replica calculation was given in 
\cite{DS1}.
However, we stress that, disregarding exceptional cases that
the asymptotic series terminates \cite{Martin,denis},
only perturbative results \cite{DS0,Drep} 
have been obtained by the replica method.

\section{CONCLUSIONS}

The QCD Dirac spectrum can be obtained from the discontinuity of the
resolvent of the Dirac operator. Its
generating function is
given by the QCD partition function with additional bosonic and
fermionic ghost quarks. Under the assumption of maximum breaking
of the axial symmetry, the low-energy limit of this generating
function can be written down on the basis of the global symmetries of
the theory. The leading infrared singularity of the resolvent, which
provides us with the slope of the average spectral density, is obtained
from a simple one-loop calculation. This result, and 
results for two other patterns of chiral symmetry breaking, can be
summarized in a single formula that depends in a natural way on the 
Dyson index of the symmetry class.

Our results for the spectral density are consistent with the infrared
singularities of the scalar susceptibility which can be calculated
by the usual chiral Lagrangian without relying on
ghost quarks. Amazingly, the two calculations also agree for $N_f =1$
when the scalar susceptibility cannot be calculated by chiral
perturbation theory. Apparently, it is possible to perform 
an analytical continuation in $N_f$. Since, as an alternative to the
supersymmetric generating function, the resolvent can also be
calculated from an analytical calculation in the number of additional
fermionic flavors, the agreement for $N_f = 1$ should not come as
a surprise.

Lattice QCD with two colors and staggered (fundamental) 
fermions is in the same symmetry class as QCD with adjoint fermions.
Our results have been extended to quenched lattice QCD 
and an impressive agreement between analytical and numerical 
results for the connected and disconnected scalar susceptibilities
has been found \cite{Berb99}. 
We are looking forward to a direct lattice calculation
of the slope of the average spectral density as well.   

\vskip 1cm
\noindent{\bf Acknowledgements}
This work was partially supported by the US DOE grant
DE-FG-88ER40388. 
One of us (D.T.) was supported in part by
``Holderbank''-Stiftung and by Janggen-P\"ohn-Stiftung.
Andrei Smilga is acknowledged for useful discussions.
The Minneapolis ITP is thanked for its hospitality.

\end{document}